\documentclass[aps,floats,amssymb,amsmath,prb,twocolumn,nofootinbib,superscriptaddress]{revtex4-1}
\usepackage{amsfonts,amssymb,amsmath}
\usepackage{graphicx}
\usepackage{amsbsy}
\usepackage{booktabs}
\usepackage{calc}
\usepackage{psfrag}
\usepackage{graphicx}
\usepackage{color}
\usepackage[dvipsnames]{xcolor}
\usepackage[unicode=true,pdfusetitle,
 bookmarks=true,bookmarksnumbered=false,bookmarksopen=false,
 breaklinks=false,pdfborder={0 0 0},backref=false,colorlinks=true]
 {hyperref}
\usepackage{geometry}
\begin{document}

%\markboth{Authors' Names}{Instructions for typing manuscripts (paper's title)}

%\catchline{}{}{}{}{}
%
%%%%%%%%%%%%%%%%%%%%%%%%%%%%%%%%%%%%%%%%%%%%%%%%%%%%%%%%%%%%%%%%%%%%

%\title{Pfaffian paired states for half-integer\\fractional quantum Hall effect\footnote{For the title, try not to
%use more than 3 lines. Typeset the title in 10 pt
%Times Roman, boldface.}
%}

\title{Pfaffian paired states for half-integer\\fractional quantum Hall effect
}

\author{\footnotesize M. V. Milovanovi\'c}

\affiliation{Scientific Computing Laboratory, Center for the Study of Complex Systems, Institute of Physics Belgrade, University of Belgrade,
Pregrevica 118\\
Belgrade, 11080,
Serbia}%\\
%milica.milovanovic@ipb.ac.rs}

\author{\footnotesize S. Djurdjevi\'c}

\affiliation{ Faculty of Natural Sciences and Mathematics, University of Montenegro, D\v zord\v za Va\v singtona bb\\
Podgorica, 81000,
Montenegro}%\\
%stevandj@ucg.ac.me}

\author{\footnotesize J. Vu\v ci\v cevi\'c}

\affiliation{Scientific Computing Laboratory, Center for the Study of Complex Systems, Institute of Physics Belgrade, University of Belgrade,
Pregrevica 118\\
Belgrade, 11080,
Serbia}%\\
%jaksa.vucicevic@ipb.ac.rs}

\author{\footnotesize L. Antoni\'c}

\affiliation{Department of Physics, Technion\\
Haifa, 32000,
Israel}%\\
%luka.antonic@campus.technion.ac.il}

%\author{\footnotesize S. Djurdjevi\'c\footnote{Typeset names in
%10~pt Times Roman. Use the footnote to indicate
%the present or permanent address of the author.}}

%\address{University of Montenegro, Faculty of Natural Sciences and Mathematics, D\v zord\v za Va\v singtona bb\\
%Podgorica, 81000,
%Montenegro\footnote{State completely without abbreviations, the
%affiliation and mailing address, including country. Typeset in 8~pt
%Times italic.}\\
%stevan.d.djurdjevic@gmail.com}

%\author{Second Author}

%\address{Group, Laboratory, Address\\
%City, State ZIP/Zone, Country\\
%second\_author@group.com}

%\begin{history}
%\received{(Day Month Year)}
%\revised{(Day Month Year)}
%\end{history}

\begin{abstract}
In this review the physics of Pfaffian paired states, in the context of fractional quantum Hall effect, is discussed using field-theoretical approaches. The Pfaffian states are prime examples of topological ($p$-wave) Cooper pairing and are characterized by non-Abelian statistics of their quasiparticles.  Here we focus on conditions for their realization and competition  among them at half-integer filling factors. Using the Dirac composite fermion description, in the presence of a mass term, we study the influence of Landau level mixing in selecting a particular Pfaffian state. While Pfaffian and anti-Pfaffian  are selected when Landau level mixing is not strong, and can be taken into account perturbatively, the PH Pfaffian state requires non-perturbative inclusion of at least two Landau levels. Our findings, for small Landau level mixing, are in accordance with numerical investigations in the literature, and call for a non-perturbative approach in the search for PH Pfaffian correlations. We demonstrated that a method based on the Chern-Simons field-theoretical approach can be used to generate characteristic interaction pseudo-potentials for Pfaffian paired states.
%The abstract should summarize the context, content
%and conclusions of the paper in less than 200 words. It should
%not contain any reference or displayed equation. Typeset the
%abstract in 8~pt Times roman with baselineskip of 10~pt, making
%an indentation of 1.5 pica on the left and right margins.
\end{abstract}

\maketitle

%\keywords{fractional quantum Hall effect; half-integer filling factor; Pfaffian paired states.}
\begin{widetext}

\section{Introduction}

The  fractional quantum Hall effect (FQHE)  \cite{fqhe} is a strongly correlated phenomenon of electrons that is observed when they are confined to two dimensions and subjected  to a strong magnetic field perpendicular to the two-dimensional plane, in which electrons live and interact. At special filling factors, i.e. ratios between the number of electrons and the number of flux quanta piercing the two-dimensional plane, experiments reveal highly entangled topological states of electrons with fractionally quantized Hall conductance, for intervals of magnetic field (or density). Almost exclusively the denominator of these fractions is an odd number, which can be traced and connected to the fermionic statistics of electrons. A surprise came when an even-denominator FQHE, at filling factor 5/2, was discovered.  \cite{will} This introduced a new paradigm in our understanding  of (even-denominator) FQHE states: they may be BCS paired states of underlying quasiparticles. If we neglect the role of spin in high magnetic fields, the most natural choice for a pairing in a fixed Landau level (LL) is the unconventional, $p$-wave pairing of spinless quasiparticles proposed in Ref.~ \onlinecite{mr}. The resulting state, Moore-Read state is also called Pfaffian due to the necessary antisymmetrization of a collection of pairs of quasiparticles - identical fermions, which do not possess any additional characteristic like spin.

The underlying quasiparticles at even-denominator fractions beside the possibility of having the BCS pairing correlations in a paired state, may in principle exist in its parent, Fermi-liquid-like (FLL) state. \cite{shs} Indeed such a state was probed and detected  at filling factor 1/2   \cite{will2}, and firstly theoretically described in Ref.~ \onlinecite{hlr}. The theoretical assessment of even-denominator FLL state(s)  may lead also to  further understanding of the physics of the BCS pairing of underlying quasiparticles. An important direction in this effort is the understanding of the FLL state that occurs at a half-integer (denominator 2) filling of the system, and, at the same time, in an artificial circumstance of a precisely half-filled LL. Namely a LL is singled out and half-filled. This mathematical limit of the physical system is highly relevant for the understanding of the real system. Our understanding of FQHE phenomena and  real circumstances of FQHE experiments call for the concept of the projection to a single LL. Very often the physics of FQHE is confined to a single LL, and we can neglect the LL mixing - the influence of other LLs. Thus if the system is at half(-integer) filling, it nearly possesses the particle-hole (PH) symmetry - the symmetry under exchange of electrons and holes that a half-filled LL has. The Halperin-Lee-Read (HLR) theory \cite{hlr}  of the FLL state at half-filling does not possess this symmetry (because it is a theory that does not include a projection to a fixed LL), but a phenomenological, effective theory with Dirac quasiparticles, proposed in Ref.~  \onlinecite{son} is manifestly invariant under exchange of electrons and holes, and describes the artificial system of electrons that is confined to a single LL.

On the other hand the Pfaffian paired state is not invariant under exchange of electrons and holes. When the PH symmetry operation is applied to the Pfaffian, a new topological state is generated, Pfaffian's conjugated partner, known as anti-Pfaffian.  \cite{ap1, ap22} Here we may ask whether a state exists, that is a collection of $p$-wave Cooper pairs and respects the PH symmetry. Indeed one may argue that the Dirac theory of the half-filled LL offers a distinct possibility  \cite{son} known as PH Pfaffian (PH symmetric Pfaffian). Before the proposal of the Dirac theory, studies that were examining possibilities of additional, negative-flux pairing, in which angular momentum of $p$-wave has opposite sign with respect to the one in Pfaffian, also proposed the PH Pfaffian. \cite{th,zf}

While the relevance of Pfaffian and especially anti-Pfaffian for the explanation of the FQHE at 5/2 is firmly established in numerical experiments  confined to a fixed LL with LL mixing (perturbatively) included via additional, three-body interactions,  \cite{rez} we do not have a support for PH Pfaffian when numerical experiments are confined to a fixed LL.  \cite{mish} But a recent experiment \cite{exp} on thermal Hall conductance is consistent with a PH Pfaffian scenario at 5/2. That the PH Pfaffian correlations and topological order may be relevant even in the absence of the PH symmetry (as is the case in  experiments) may be shown by careful examination of various experimental probes as discussed in Ref.~ \onlinecite{mafe}.

Thus the question is whether for sufficiently strong LL mixing, that cannot be treated perturbatively (as it is done in all numerical experiments confined to a single LL), we can reach a regime in a uniform system when PH Pfaffian correlations prevail. Or, is disorder  needed to install the effective PH Pfaffian correlations? \cite{ph1, ph2}   In any case LL mixing may play decisive role in selecting a specific kind of Pfaffian state in experiments. In the following sections we will review our work 
 \cite{ avm} that used Dirac and Chern-Simons (CS) field-theoretical description to examine the role of LL mixing and explore pairing at half-integer fillings, in general.

In Section II we will review the  Dirac theory of the FLL state of underlying quasiparticles - composite fermions at a  half-filled LL, and select and describe a version of the theory that is best fitted for a  description of Pfaffian paired states. The mass term  in this theory mimics LL mixing (for small LL mixing has the role of those additional (three-body) interactions in the electron representation),  and the limiting behavior of large mass may be identified  with the usual HLR picture of the FLL state of FQHE at half-filling.

In Section III within this version of the  Dirac theory, we will probe the question of topological pairing instabilities in a mean-field approximation (as usual in topological explorations when we assume that topological characterization is immune to the neglect of fluctuations). Instabilities will originate from the minimal coupling term i.e. the coupling with the CS gauge field, and we will be disregarding the remaining influence of the Coulomb interaction which has a pair-breaking effect. Our interest will be to find which kind of Pfaffian will prevail at certain LL mixing, if we assume a pairing instability.

In Section IV we will discuss which model Hamiltonians for electrons i.e. effective interaction pseudo-potentials (PPs)  in fixed LLs lead to Pfaffian states. Using CS field-theoretical description we recover dominant, already known PPs for Pfaffian and anti-Pfaffian in a fixed LL, and discuss the necessity to include non-perturbatively at least one more LL to establish PH Pfaffian correlations, and list pertinent PPs.   \cite{ djm}  Section V is reserved for a discussion and conclusions.

\section{Theoretical approaches to the physics at a half-integer filling}

\subsection{Wave-function approach}

The basic explanation of the FQHE rests on the Laughlin wave function - the ground state wave function for the most prominent effect at filling 1/3.
 \cite{laugh1} The wave function captures the basic correlations of electrons in a constrained space of an isolated LL. To introduce the Laughlin wave function we start with the single-particle Hamiltonian,
\begin{equation}
H = \frac{ (\bold{p} - \bold{A})^2 }{2 m_{\rm e}} ,
\label{spH}
\end{equation}
of a particle in a constant magnetic field, $\bold{B} = B  \mathbf{z} $, with  $ A_{x} = - (B/2) y $ and $A_{y} = (B/2)  x $, in a rotationally symmetric gauge. We fixed $c = 1$, $e = 1$, and $\hbar = 1$. The physics of FQHE is largely confined to a fixed LL and in the case of filling factor 1/3, to the lowest LL (LLL). In the rotationally symmetric gauge and in the LLL, the appropriate basis is given by the following single particle wave functions, 
\begin{equation}
  \Psi_n ({\bf{r}}) =  \frac{1}{\sqrt{2 \pi l_B^{2+2n} 2^n n!}} z^n \exp\{-(\frac{1}{4 l_B^2}) |z|^2\},
\label{spwf}
\end{equation}
where $l_B = \sqrt{\frac{\hbar c}{e B}}$, and $ n = 0, 1, 2 \ldots $ is the guiding center angular momentum number. Apart from the exponential factor, these wave functions depend only on the coordinate $z = x + i y$, i.e. they make a holomorphic description, when we neglect the factor which is the same for each 
$  \Psi_n ({\bf{r}}) $. Thus many-body wave functions of frozen spin electrons become polynomials in the $z$ coordinate(s) in the LLL,  as in the following expression,
\begin{equation}
\Psi ({\bf r}_1 , {\bf r}_2 , \ldots , {\bf r}_{N_e} ) = P( z_1 , z_2 , \ldots , z_N ) \exp\{-(\frac{1}{4 l_B^2}) \sum_{i=1}^{N_e} |z_i|^2\}.
\label{poly}
\end{equation}
The Laughlin wave function at filling factor 1/3 is specified by the Laughlin-Jastrow choice for $P$,
\begin{equation}
 P_{L-J}  ( z_1 , z_2 , \ldots , z_{N_e} ) = \prod_{i<j} (z_i - z_j)^m ,
\label{lch}
\end{equation}
with $m = 3$. In this polynomial the highest power of any $z_i ;\; i = 1,2, \ldots , N_e $ is $N_m  = m (N_e  - 1)$ and this number also specifies the number of (single-particle) states available to the system i.e. the number of flux-quanta piercing the system, $N_\phi = N_m  + 1$. Thus the ratio $ N_e /N_\phi $ becomes 1/3 in the thermodynamic limit when $m = 3$. 

For monotonically decreasing with distance repulsive interactions like Coulomb, we may expect an extreme capacity of the wave-function to minimize the interaction energy. Namely, as a function of a fixed electron coordinate, the wave function has all ($N_m $)
zeros  on the other electrons, $m = 3$ per electron, though only one zero is required by Fermi statistics.  Equivalently, we may  say that the zero on any other electron is of the $m^{\rm th}$  order as we study the limiting behavior when a fixed electron approaches any other  in (\ref{lch}).

Following the same logic, we may attempt the same construction at filling factor 1/2,  but, because  $m = 2$ in (\ref{lch}), in this case, we need additional factors that will ensure that the wave function is antisymmetric. These additional factors should not contribute or change the value of $N_m $ in the thermodynamic limit ($ m N $), and thus, as additional factors in the total wave function, may be considered as its ``neutral part" - the part that does not see the macroscopic flux. (The Laughlin-Jastrow part  (\ref{lch}) would represent the charged part.)  The neutral part may describe a collection of fermionic quasiparticles (that do not see any macroscopic flux i.e. external magnetic field), and they may be  in the first approximation non-interacting (make a FLL - state), or they may come in BCS paires (make a bosonic condenaste and possibly a gapped state). Indeed experiment and theory are equivocal that the state at filling factor 1/2 (in GaAs structures) is a FLL state of underlying quasiparticles, and the state at filling factor 5/2 (in GaAs  \cite{will}) is effectively a gapped state of half-filled second LL of frozen-spin (spinless) electrons, in which quasiparticles may pair. The exact topological nature of the paired state at filling factor 5/2 is still under debate.

But we may say that the most theoretically appealing (the most simple and natural BCS pairing) guess for the gapped state at the half-integer filling factors (in 
various experimental set-ups) is proposed in Ref.~\onlinecite{mr}, and goes under name Moore-Read state or Pfaffian (state). The Pfaffian wave function in the LLL is
\begin{equation}
\Psi_{\rm Pf} =   \sum_\sigma {\rm sgn} \; \sigma \{ \frac{1}{(z_{\sigma(1)} - z_{\sigma(2)})}  \cdots
\frac{1}{(z_{\sigma(N_e -1)} - z_{\sigma(N_e )})}   \} \prod_{k<l} (z_{k} - z_{l})^2 ,
 \label{pf}
\end{equation}
where the sum is over all permutations of $N_e$ objects where $N_e$ is an even number. We omitted the exponential factors and the expression is unnormalized. In mathematics, if $A = \{ a_{ij} \}$ is $ N \times N$ anti-symmetric matrix, and $N$ is even, its Pfaffian is 
\begin{equation}
{\rm pf}(a_{ij}) = {\rm pf}(A) = \frac{1}{2^{N/2} (N/2)!}  \sum_{\sigma \in S_N}  {\rm sgn} \; \sigma \prod_{i = 1}^{N/2} a_{\sigma(2 i - 1) \sigma(2i)},
\label{pff}
\end{equation}
and  $ {\rm pf}(A)^2 = {\rm det}(A)$. In more physical terms we see that the sum in the Moore-Read wave function describes the antisymmetrization of a collection of Cooper pairs, where each pair wave function, $g({\bf{r}})$, where ${\bf{r}}$ is the relative coordinate of a pair, can be described as 
\begin{equation}
g({\bf{r}}) \sim \frac{1}{z}.
\label{Cpwf}
\end{equation}
This special algebraic decay is the hallmark of the Pfaffian (Moore-Read) wave function, and expresses a special kind of topological, long-range entanglement in this function that represents a $p$-wave pairing.  The construction is given in the LLL, but can be easily generalized and considered in the second LL, i.e. in any isolated LL.

The highest power of any $z_i $ in the Pfaffian wave function is $N_m = 2 N_e - 3$ i.e. $N_m = 2 N_e -{\cal S}$,  where ${\cal S} = 3$ is so-called shift - a topological number that characterizes a state of a FQHE system on a curved background, such as a sphere. If a state is PH symmetric, the shift should be invariant under the PH exchange. We require $N_e + N_h = N_m + 1$,  i.e. the number of electrons, $N_e$, plus the number of holes, $N_h$, should be equal to the number of available single-particle states. Thus the state that we get by applying the PH transformation on Pfaffian, is a distinct state, anti-Pfaffian, with shift equal to $-1$. This anti-Pfaffian state, that has distinct topological features with respect to Pfaffian, was firstly described in Refs. \onlinecite{ap1} and \onlinecite{ap22}.

We may wonder whether we may still have a $p$-wave pairing (the smallest angular momentum pairing of spinless electrons) in a many-body wave function that is invariant under PH exchange. It is not hard to see that in this case we must have $N_m = 2 N_e - 1$, and this implies some  kind of a microscopic negative flux or simply reversed $p$-wave pairing as in
\begin{equation}
g_{\rm ph}({\bf{r}}) \sim \frac{1}{z^*}.
\label{phCpwf}
\end{equation}
The naive guess would be that by doing the projection to the LLL, in the first approximation, we have
\begin{equation}
g_{\rm ph}({\bf{r}}) \sim z .
\label{PphCpwf}
\end{equation}
But, because for any set of complex numbers $z_i$, $  i = 1, 2, \ldots, N,   ; N$ even, and $N > 2$,
\begin{equation}
{\rm pf}(z_i - z_j ) = 0,
\end{equation}
this does not lead to a non-trivial state in the LLL. Thus the question is whether a half-filled isolated LL with special interactions can support a gapped state with PH symmetry, i.e. PH (symmetric) Pfaffian. In the case of Pfaffian and anti-Pfaffian special interactions exist in an isolated LL  \cite{rh} (and they do not respect the PH symmetry). Furthermore, the negative flux pairing expression in (\ref{phCpwf}) calls for inclusion of other LLs, and maybe only with significant LL mixing, when the PH symmetry is broken, we can stabilize the pairing correlations in  (\ref{phCpwf}). Even in this case we will call this exotic state PH Pfaffian.

\subsection{Field-theoretical approach}

\subsubsection{Quasiparticles in the FQHE and the HLR theory at half-filling}

We may separate the phase part from the rest of the Laughlin wave function at filling factor $1/m$, where $m = 3$, or from the Laughlin-Jastrow part of a ground state wave function at half-filling, when $m = 2$, and, then, define a decomposition into two parts of any many-electron wave-function, $ \Psi_{\rm e} $,  as
\begin{equation}
\Psi_{\rm e}   ({\bf r}_1 , {\bf r}_2 , \ldots , {\bf r}_{N_e} )  = \prod_{i<j} \frac{(z_i - z_j)^m}{|z_i - z_j |^m} \Psi_{\rm qp}  ({\bf r}_1 , {\bf r}_2 , \ldots , {\bf r}_{N_e} ) .
\label{cstr}
\end{equation}
The wave function  $\Psi_{\rm qp}  ({\bf r}_1 , {\bf r}_2 , \ldots , {\bf r}_{N_e} ) $ represents a wave function of quasiparticles after the unitary transformation defined by the phase factor: in the Laughlin  $(m = 3)$ case quasiparticles are bosons, and at half-filling  $(m = 2)$ they are fermions. This defines a Chern-Simons transformation, or what we will refer to as a Zhang's construction of quasiparticles.  \cite{zhang} In the field-theoretical terms quasiparticles induce field ${\bf a}$ - they are the sources of an artificial (internal) magnetic field $b$ that also acts as an additional field on quasiparticles,
\begin{equation}
\rho_{\rm qp} = - \frac{1}{m} \frac{\nabla \times {\bf a}}{2 \pi} = - \frac{1}{m} b .
\label{cseq}
\end{equation}
In (\ref{cseq})  $\rho_{\rm qp}$ is the quasiparticle density. We will discuss the Chern-Simons field-theoretical approach to the system at half-filling, i.e. the HLR theory with more mathematical details below. Here we will note that in a mean-field picture the internal field will cancel the external field. As a first approximation to the half-filling problem we will find that the ground state  in the quasiparticle representation is simply a Slater-determinant of free waves that are filling a Fermi sphere in the inverse space in two dimensions, i.e. it represents a gas of fermionic quasiparticles. (The amplitude part of the Laughlin-Jastrow factor can be recovered in the  field-theoretical approach by the RPA treatment of the density harmonic fluctuations.)

Therefore, in the Zhang's quasiparticle construction to each electron at position $w$ is attached the following phase factor:
\begin{equation}
 \prod_{i} \frac{|z_i - w |^m}{(z_i - w)^m} ,
\label{csphasefactor}
\end{equation}
a flux tube. The ensuing quasiparticle sees two gauge fields: external and internal - it is a quasiparticle that possesses charge, and the density of quasiparticles  is equal to the density of electrons.

On the other hand in the Read's construction  \cite{read} of quasiparticles we start with the notion of fluxes (flux quanta or vortices) that can be introduced by external field in the system, and can be described by the following construction,
\begin{equation}
 \prod_{i} (z_i - w)^m  ,
\label{Readfactor}
\end{equation}
i.e. by insertion of $m$ Laughlin quasiholes. We can make this object neutral by adding a unit of charge, more precisely an electron, to it, and in this way define the Read's quasiparticles as neutral objects, number of which is propotional to the number of external field flux quanta  piercing the system. This view is in a way a dual approach  (equivalent description of the same theory from a different point of view)  that was initially applied to bosonic systems where the description in terms of elementary particles - bosons was traded for the description in terms of excitations - vortices.\cite{lf}

In any case both approaches take into account the precise commensuration between the number of electrons and the number of flux quanta in a system at a fixed filling factor, in our case 1/2.

The Chern-Simons approach at 1/2, based on the Zhang's construction of quasiparticles, begins with the following Lagrangian (density),
\begin{equation}
{\cal L} = \Psi_{\rm cf}^{*} (i  \partial_t - A_0 - a_0 ) \Psi_{\rm cf} - \frac{\Psi_{\rm cf}^{*} ({\bf p} - {\bf A} - {\bf a} )^2 \Psi_{\rm cf}}{2 m} 
- \frac{1}{2} \frac{1}{4 \pi} a \partial a .
\label{hlrL}
\end{equation}
In (\ref{hlrL}), $ \Psi_{\rm cf}$ represents a fermionic (Grassmann quasiparticle) field, and  the Chern-Simons term is defined by $ a \partial a \equiv \epsilon^{\mu \nu \lambda} a_{\mu} \partial_{\nu} a_{\lambda} $, $\mu, \nu, \lambda = 0, 1, 2 $ (denote one time and two spatial coordinates), the summation over repeated indecies is understood, and $a_\mu = (a_0 , {\mathbf a})$  is a three-vector. The ${\rm cf}$ stands for composite fermions, a general name for underlying quasiparticles. 

Considering the classical equations of motion, from $ \frac{\delta {\cal L}}{\delta a_0 } = 0 $, we get
\begin{equation}
- \Psi_{\rm cf}^{*} \Psi_{\rm cf}  -  \frac{1}{2} \frac{\nabla \times {\bf a}}{2 \pi} = 0.
\end{equation}
(Above  $\nabla \times {\bf a}$ denotes the $z$ component of the vector, and can be considered as a scalar in this two-dimensional theory.)
In the mean-field, when we assume that the density of quasiparticles is uniform, the internal field, $ \frac{\nabla \times {\bf a}}{2 \pi} $, exactly cancels the uniform external field at half-filling,
\begin{equation}
  \frac{\nabla \times {\bf A}}{2 \pi} = 2  \overline{ \Psi_{\rm e}^* \Psi_{\rm e} } =  2  \overline{\Psi_{\rm cf}^* \Psi_{\rm cf} },
\end{equation}
where  $  \overline{ \Psi_{\rm e}^* \Psi_{\rm e} } $ stands for the uniform electron density.

The Lagrangian in (\ref{hlrL}) is the basis or starting point for the HLR theory, which describes the physics at 1/2 as a FLL state of (fermionic) quasiparticles. We may notice, from the form of the Lagrangian, that the electron density-current vector is equal to the one of quasiparticles,
\begin{equation}
- \frac{\delta {\cal L}}{\delta A^\mu } = j_{\rm e}^{\mu} = j_{\rm cf}^{\mu}.
\end{equation}

\subsubsection{Dirac quasiparticle description of half-filled Landau level and at half-filling}

In this section we will first review the Dirac theory for a half-filled LL proposed in  Ref.~\onlinecite{son}   and then consider its extension in the presence of a mass term that is relevant for the general case (with LL mixing) at half-filling. 

We start with an isolated LL (of classical electrons) that is half-filled. It has the PH symmetry - the symmetry under exchange of electrons and holes. The low-energy physics of a zeroth LL of Dirac electrons in the weak coupling limit should correspond to the low-energy physics of isolated LL (of classical electrons).\cite{son}
Thus we consider the Dirac problem in an external (magnetic) field, which is a background field (no dynamics):
\begin{equation}
{\cal L}_D = i {\overline \Psi} \gamma^\mu D_{\mu}^{A} \Psi +  {\rm interactions} = i {\overline \Psi} (\gamma^0 D_t + \boldsymbol{\gamma}\cdot{\bf D} ) \Psi + {\rm interactions}
\end{equation}
where $ D_t = \frac{\partial}{\partial t} + i A_0  $ and ${\bf D} = {\bf \nabla} - i {\bf A}$, and $\gamma^\mu , \;\mu = 0, 1, 2 $ are $2 \times 2$ gamma matrices for the  Dirac description in  two spacial dimensions, and  $ \Psi $ is a two-component Grassmann field.

The Dirac system is a neutral system and there is no Hall conductance. To make up for this i.e. to continue to discuss an isolated LL (of classical electrons),
which has $1/(4 \pi)$  of the units $(e^2/\hbar)$ of Hall conductance, we consider
\begin{equation}
{\cal L}_A = i {\overline \Psi} \gamma^\mu D_{\mu}^{A}  \Psi - \frac{A \partial A}{8 \pi} + {\rm interactions}.
\end{equation}
If we define the density-current of electrons as
\begin{equation}
j^\mu_{\rm el} = - \frac{\delta {\cal L}}{\delta A^\mu},
\end{equation}
it follows that for densities,
\begin{equation}
\rho_{\rm el} = \rho_{\rm D} + \frac{{\bf \nabla} \times {\bf A}}{4 \pi}.
\end{equation}
Because, ${\overline \rho}_{\rm D}$ (average density of the Dirac system) = 0, we have a non-zero density of electrons 
\begin{equation}
\frac{\overline{\rho}_{\rm el}}{B} = \frac{1}{2},
\end{equation}
where $ B = \frac{{\bf \nabla} \times {\bf A}}{2 \pi} $ is the uniform external magnetic field. Also
\begin{equation}
{\bf j}_{\rm el} = {\bf j}_{\rm D} + \hat{\epsilon} \frac{{\bf E}}{4 \pi} ,
\end{equation}
where $\hat{\epsilon}$ is a $2 \times 2$ matrix, $ \epsilon_{xy} = - \epsilon_{yx} = 1$, $\epsilon_{xx} = \epsilon_{yy} = 0$. Thus, with $\overline{\rho}_{\rm D} = 0 $ and $ {\bf j}_{\rm D} = 0$,  we are at half-filling, and the Hall conductance is equal to $ \frac{1}{4\pi } (\frac{e^2}{\hbar})$.

Following Ref.   \onlinecite{son}, in a dual picture,  we postulate a new Lagrangian, $ {\cal L}$,  with new - dual Dirac field $\chi$ :
\begin{equation}
{\cal L} = i \overline{\chi} \gamma^\mu D_{\mu}^{a} \chi + a \frac{\partial A}{4 \pi} - \frac{A \partial A}{8 \pi} + \cdots
\end{equation}
where $\cdots$ denotes higher order terms. (We will ignore these higher order terms below and consider classical equations of motion in the framework of the linear response theory.) Why would we expect this Lagrangian in a dual picture? We provide an analysis with more details below, but here we may note that the Dirac (two-component) formalism is expected also in a dual picture, because it makes possible that the PH symmetry is manifestly included as demonstrated in Ref. \onlinecite{son}. Also note that the dual fermion is not directly coupled  to the external field, and, as we show below, the Lagrangian describes a Dirac system at a finite density,  in agreement with our expectation that  the system is in a FLL state of quasiparticles.   For further details on the dual approach see  Refs.~ \onlinecite{sw} and \onlinecite{ss}.

\begin{itemize}%[(ii)]

\item It seems that $\chi$'s represent Read's quasiparticles. Indeed, if we consider the following equation of motion,
\begin{equation}
0 = \frac{\delta {\cal L}}{\delta a_0} = - \rho_{\chi} +  \frac{{\bf \nabla} \times {\bf A}}{4 \pi},
\end{equation}
we can conclude that the density of $\chi$ depends on the number of flux quanta. On the other hand,
\begin{equation}
\rho_{\rm el}  = - \frac{\delta {\cal L}}{\delta A_0} = -  \frac{{\bf \nabla} \times {\bf a}}{4 \pi} +  \frac{{\bf \nabla} \times {\bf A}}{4 \pi},
\end{equation}
and, at half-filling, in the mean-field approximation, ${\bf \nabla} \times {\bf a} = 0$. Thus, $\chi$'s do not experience any uniform, non-zero gauge field,
$b =  \frac{{\bf \nabla} \times {\bf a}}{2 \pi}$, that couples $\chi$'s indirectly to the external field. Therefore, $\chi$'s are, in the first approximation, neutral objects, but with the Dirac's singularity in the inverse space at ${\bf k} = 0$. In this way they have a non-analytical feature that we do not expect from a description that is based on Read's quasiparticles.
% There are other suggestions  \cite{js1,js2} which are based on a uniform Berry curvature description (no Dirac singularity).
 We find that the effective theories based on the description with the Dirac's quasiparticle are very useful when considering the pairing physics, as they capture the time-reversal and parity breaking (that is essential for the pairing physics) as we will explain later in this section.

\item We expect that the effective theory of a half-filled LL should describe a Fermi-liquid of quasiparticles (if we do not consider the BCS instability). Indeed, in the mean-field approximation, in the first approximation, the internal field $(b)$ is zero, and the theory describes a Dirac Fermi-liquid.

\item If we vary ${\bf a}$ in ${\cal L}$ we find
\begin{equation}
{\bf j}_{\rm D} =  \hat{\epsilon} \frac{{\bf E}}{4 \pi} .
\label{firsteq}
\end{equation}
Also,
\begin{equation}
{\bf j}_{\rm el} =  - \frac{\delta {\cal L}}{\delta {\bf A}} =  \hat{\epsilon} \frac{{\bf E} - {\bf e}}{4 \pi} ,
\label{secondeq}
\end{equation}
where ${\bf e}$ is the electric field due to the potential $ a^\mu$. Next, we assume that even in the presence of disorder, the PH symmetry is respected, and in the linear response we have,
\begin{equation}
{\bf j}_{\rm D} = \hat{\sigma}^{\rm D} {\bf e},
\label{thirdeq}
\end{equation}
where $\sigma_{xx}^{\rm D} = \sigma_{yy}^{\rm D} \neq 0 $ represents a longitudinal conductance, and  $\sigma_{xy}^{\rm D} = \sigma_{yx}^{\rm D} = 0 $ (the Hall conductance is zero). The zero Hall conductance is an expression of the PH symmetry and a property of Dirac fermions. These three equations, (\ref{firsteq}),
(\ref{secondeq}), and (\ref{thirdeq}),  combined lead to the conclusion that the Hall conductance of electrons is $ \frac{1}{2} (\frac{e^2}{h}) $, which we expect to be the case in the theory of the system with classical electrons that respects the PH symmetry.\cite{pot} 
\end{itemize}

It is important to notice that  $\sigma_{xy}^{\rm D} = \sigma_{yx}^{\rm D} = 0 $  is not an only natural ``choice" for the response of the non-interacting Dirac system (conus) to a 
perturbation due to a gauge (internal $a^\mu $) field. 
To get  the Hall conductance  we assume the presence of the mass term in the non-interacting Dirac description,
\begin{equation}
{\cal L}_{\rm D}  = i \overline{\chi} \gamma^\mu D_{\mu}^{a} \chi - m \overline{\chi} \chi .
\end{equation}
The  $\sigma_{xy}^{\rm D}  $  can be found by integration of Berry curvature in the inverse (${\bf k}$) space,  \cite{pot,niu} by choosing a specific gauge for eigenstates, and integrating over occupied states. In this way we can get contributions (in units  $ e^2/\hbar $):
\begin{equation}
sgn(m) \frac{1}{4 \pi} (1 - \frac{|m|}{\sqrt{k_F^2 + m^2}}),
\end{equation}
from the positive-energy states that are filled for $ 0 \leq |{\bf k}| < k_F $, and
\begin{equation}
- sgn(m) \frac{1}{4 \pi} ,
\end{equation}
from the negative-energy states. 
 There are two natural ways to take into account these two contributions: (1) to add them,
\begin{equation}
\sigma_{xy}^{\rm D}  = - \frac{m}{\sqrt{k_f^2 + m^2}} ,
\end{equation}
i.e. adopt a ``dimensional regularization", or (2) to consider only the contribution from the positive energy solutions:
\begin{equation}
\sigma_{xy}^{\rm D}  = sgn(m) \frac{1}{4 \pi} (1 - \frac{|m|}{\sqrt{k_F^2 + m^2}}) ,
\label{pv}
\end{equation}
i.e. adopt a ``Pauli-Villars regularization".  It is obvious that in order to get an appropriate response in the Dirac theory (of the half-filled LL) we need to assume and apply the dimensional regularization in the field-theoretical treatment.

We can also conclude that by choosing an appropriate singular gauge (phase) transformation on the negative energy eigenstates, we can switch from the dimensional regularization to the Pauli-Villars regularzation (and vice versa). This transformation can be understood as an adoption of a new quasiparticle picture and a new Lagrangian (here without higher order terms) :
\begin{equation}
{\cal L} = i \overline{\chi}^{\rm qp} \gamma^\mu  D^a_\mu \chi^{\rm qp} - \frac{a \partial a}{8 \pi}   + a \frac{\partial A}{4 \pi} - \frac{A \partial A}{8 \pi} .
\label{qplag}
\end{equation}
To find the same response as before we have to adopt Pauli-Villars regularization (when integrating out fermions and generating quadratic terms in $a$) with a positive mass to cancel the second term in ${\cal L}$. Physically we indeed switched to a new quasiparticle picture of Zhang's type. To see that let's consider the full theory with a positive $(m > 0)$ mass term:
\begin{equation}
{\cal L} = i \overline{\chi}^{\rm qp} D_a \chi^{\rm qp} - m \overline{\chi}^{\rm qp} \chi^{\rm qp}   - \frac{a \partial a}{8 \pi}   + a \frac{\partial A}{4 \pi} - \frac{A \partial A}{8 \pi} .
\label{PVL}
\end{equation}
\begin{itemize}%[(ii)]

\item From the equations of motion,
\begin{equation}
0 = \frac{\delta {\cal L}}{\delta a^\mu } = - j^{{\rm  qp}, \mu}_{\chi} - \frac{\partial a}{4 \pi} + \frac{\partial A}{4 \pi},
\label{eqm1}
\end{equation}
and
\begin{equation}
j_{\rm el}^\mu = -  \frac{\delta {\cal L}}{\delta A^\mu } =  \frac{\partial A}{4 \pi} -  \frac{\partial a}{4 \pi},
\label{eqm2}
\end{equation}
it follows that, $ j_{\rm el}^\mu = j^{{\rm qp}, \mu}_{\chi} $, as usual in the Chern-Simons theory, i.e. the theory directly relates to the Zhang's quasiparticle construction, and 
\item  if we let $ m \rightarrow \infty $ the effective Lagrangian becomes the HLR after the shift $ a^\mu \rightarrow a^\mu + A^\mu $. \cite{rev}
\end{itemize}
We can conclude that the Lagrangian in (\ref{PVL}), with $m = 0$,  describes the physics of an isolated (PH symmetric) LL  using the Zhang's quasiparticle picture. The introduction of non-zero  $m$ represents LL mixing, i.e. a measure of the inclusion of other LLs, so that for large $m$ we can recover the HLR theory that does not reduce the effective physics of the electron system to a single LL.

\section{Pfaffian paired states at half-integer filling}

In this section we will adopt the Dirac quasiparticle picture that is given by the Lagrangian in (\ref{PVL}) for a FQHE system at a half-integer filling factor. Thus the starting Lagrangian is
\begin{equation}
{\cal L} = i \overline{\chi} \gamma^\mu D^a_\mu \chi - m \overline{\chi} \chi   - \frac{m}{|m|}  \frac{a \partial a}{8 \pi}   + a \frac{\partial A}{4 \pi} - \frac{A \partial A}{8 \pi} ,
\label{PVLGen}
\end{equation}
where for simplicity we omitted qp letters when writing $\chi$ fields with respect to (\ref{PVL}), but we should be aware that for any probes (perturbative expansions) the Pauli-Villars regularization is understood. We generalized the Lagrangian in (\ref{PVL}) for both signs of mass $m$ (to cancel the additional contribution due to the assumed Pauli-Villars regularization, the first term in (\ref{pv})). It follows that
\begin{equation}
 j^{ \mu}_{\chi} =  - \frac{m}{|m|}  \frac{\partial a}{4 \pi} + \frac{\partial A}{4 \pi},
\label{feq}
\end{equation}
and 
\begin{equation}
 j^{ \mu}_{\rm el} =  -   \frac{\partial a}{4 \pi} + \frac{\partial A}{4 \pi},
\label{seq}
\end{equation}
as a generalization of (\ref{eqm1}) and (\ref{eqm2}) to both signs of mass.
Exactly at half-filling, i.e. when in a uniform, constant magnetic field we have on average one electron per two flux quanta, we may solve (\ref{feq}) in the Coulomb gauge, $ {\bf \nabla} \cdot {\bf a} = 0 $. The solutions are \cite{zhang}
\begin{equation}
a_x ({\bf r}) = 2 \frac{m}{|m|}  \int d{\bf r}' i \frac{y - y'}{|{\bf r} - {\bf r}'|^2} \delta \rho_\chi ({\bf r}'), \label{ax}
\end{equation}
and
\begin{equation}
a_y  ({\bf r}) = -  2 \frac{m}{|m|}  \int d{\bf r}' i \frac{x - x'}{|{\bf r} - {\bf r}'|^2} \delta \rho_\chi ({\bf r}'), \label{ay}
\end{equation}
and $ \delta \rho_\chi ({\bf r}') = \chi^\dagger ({\bf r}')  \chi ({\bf r}') - \bar{\rho}$, where $\bar{\rho}$ is a constant (external flux density). 
We would like to analyze the effect on pairing of the  interaction term,
\begin{equation}
V_{\rm int} = - {\bf a} {\overline \chi} \boldsymbol{\gamma} \chi .
\end{equation}
%(We use the relativistic expansion $ - a_\mu \gamma^\mu = - a_0 \gamma^0 + {\bf a} {\bf \gamma} $, and the extra minus sign comes from 
%interpreting the $ {\bf a} {\bf \gamma} $ as an interaction term in the Lagrangian.)
 In the following representation of $\gamma $ matrices,
\begin{equation}
\gamma^0 = \sigma_3 , \,\,\, \gamma^1 = i \sigma_2 , \,\,\, \gamma^2 = - i \sigma_1 ,
\end{equation}
where $ \sigma_i , i = 1, 2, 3 $ are Pauli matrices, we have
\begin{equation}
V_{\rm int} = - {\bf a}  \chi^{+}  \boldsymbol{\sigma} \chi .
\end{equation}
In this representation we have the following expression for the  interaction:
\begin{eqnarray}
V_{\rm int} &=& - i 2 \frac{m}{|m|}  \int d{\bf r}' \delta \rho_\chi ({\bf r}')  \chi^\dagger ({\bf r}) \left[
  \begin{array}{cc}
   0 &  \frac{\bar{z} - \bar{z}'}{|{\bf r} - {\bf r}'|^2} \\
   - \frac{z - z'}{|{\bf r} - {\bf r}'|^2} & 0 \\
  \end{array}
\right] \chi({\bf r}).
\end{eqnarray}
%and $ \delta \rho_\chi ({\bf r}') = \chi^\dagger ({\bf r}')  \chi ({\bf r}') - \bar{\rho}$, where $\bar{\rho}$ is a constant (external flux density).
% The constant part gives no contribution to $V_{st}$.
On the other hand the presence of the mass term  in the Dirac system leads to the following eigenproblem,
\begin{equation}
 \left[
  \begin{array}{cc}
   m - \epsilon & k_- \\
   k_+ & - m - \epsilon \\
  \end{array}
\right] \chi({\bf k}) = 0,
\end{equation}
where $k_- = k_x  - i  k_y$ and  $k_+ = k_x  + i k_y$. The   positive eigenvalue, $ \epsilon = \sqrt{|{\bf k}|^2 + m^2} \equiv E_{\bf k}$, corresponds  to the following eigenstate,
\begin{equation}
\chi_E = \left[
  \begin{array}{c}
   m + E_{\bf k} \\
   k_+ \\
  \end{array}
\right]  \frac{1}{\sqrt{2 E_{\bf k} (E_{\bf k} + m)}}.
\label{eigenD}
\end{equation}
As we consider relevant only (positive energy) states around $k_F$, we will keep only these states in the expansion over ${\bf k}$-eigenstates of field $\chi({\bf r})$, and, further, only consider the BCS pairing channel in $V_{\rm int}$.
Thus (in the second-quantized notation)
\begin{equation}
\chi({\bf r}) = \frac{1}{\sqrt{2V}} \sum_{{\bf k}} \exp\{i {\bf k} \cdot {\bf r}\} \chi_E ({\bf k}) a_{\bf k} + \cdots ,
\label{exp}
\end{equation}
and
\begin{eqnarray}
V_{\rm int}^{\rm BCS} & = & \frac{m}{|m|} \frac{2 \pi}{8 V} \sum_{{\bf k},{\bf p}} a_{{\bf k}}^\dagger a_{{\bf p}} a_{-{\bf k}}^\dagger a_{-{\bf p}}
 \nonumber \\
& \times & \frac{1}{E_{{\bf k}} E_{{\bf p}} (m + E_{{\bf k}}) (m + E_{{\bf p}}) }  \nonumber \\
& \times & \{(m + E_k) (m + E_p) + k_- p_+ \} \nonumber \\
& \times &\left[
  \begin{array}{cc}
   m + E_k ,  & -  k_-
  \end{array}
\right]
\left[
  \begin{array}{cc}
   0 & \frac{1}{k_+ - p_+} \\
 - \frac{1}{k_- - p_-}  & 0 \\
  \end{array}
\right]
\left[
  \begin{array}{c}
   m + E_p \\
  -  p_+ \\
  \end{array}
\right].
\nonumber \\
\label{maineq}
\end{eqnarray}
We used:
$ \int d{\bf r} \frac{1}{z} \exp\{i {\bf k}{\bf r}\} = i \frac{2 \pi}{k_+}.$
%The Cooper channel can be cast in the
%following form
%\begin{eqnarray}
%V_{st}^{BCS} & = & \frac{2 \pi}{8 V }  \sum_{{\bf k},{\bf p}} a_{{\bf k}}^\dagger a_{{\bf p}} a_{-{\bf k}}^\dagger a_{-{\bf p}} \frac{1}{E_{{\bf k}} \cdot E_{{\bf p}} }
% \nonumber \\
% & \times & \{- 4 |m||{\bf k}| |{\bf p}| \frac{i \sin (\theta_p - \theta_k )}{|{\bf k} - {\bf p}|^2 } \nonumber \\
% && - \frac{m}{|m|} (E_k + E_p + 2 m) (E_k - m)(E_p - m)  \times \frac{i \sin 2(\theta_p - \theta_k )}{|{\bf k} - {\bf p}|^2 }  \nonumber \\
%&& + 4 |m||{\bf k}| |{\bf p}| \frac{(\lambda - 1 )}{|{\bf k} - {\bf p}|^2 } \nonumber \\
%&& - 4 |m||{\bf k}| |{\bf p}| \frac{ \cos (\theta_p - \theta_k ) - 1}{|{\bf k} - {\bf p}|^2 } \nonumber \\
%&& - \frac{m}{|m|} (E_k + E_p + 2 m) (E_k - m)(E_p - m)  \times \frac{ \cos 2(\theta_p - \theta_k ) - 1}{|{\bf k} - {\bf p}|^2 }  \},
 %\label{BSCeff}
%\end{eqnarray}
%where $ \lambda = \frac{|{\bf k}|^2 + |{\bf p}|^2}{2 |{\bf k}||{\bf p}|}$.
 We may rewrite this expression (taking into account the antisymmetry of the
fermionic operators) as
\begin{equation}
V_{\rm int}^{\rm BCS}  =   \sum_{{\bf k},{\bf p}}    V_{{\bf k}{\bf p}}      a_{{\bf k}}^\dagger a_{{\bf p}} a_{-{\bf k}}^\dagger a_{-{\bf p}} ,
\end{equation}
where
\begin{eqnarray}
V_{{\bf k}{\bf p}} & = & \frac{2 \pi}{8 V } \frac{1}{E_k \cdot E_p } \times
 \nonumber \\
 & \times & \Bigg[- 4 |m|kp \frac{i \sin (\theta_p - \theta_k )}{|{\bf k} - {\bf p}|^2 } \nonumber \\
 && \;\;- \frac{m}{|m|} (E_k + E_p + 2 m) (E_k - m)(E_p - m) \times \frac{ \exp\{ i 2(\theta_p - \theta_k )\} - 1}{|{\bf k} - {\bf p}|^2 } \Bigg].
 \label{Vkp}
\end{eqnarray}
Now we will adopt the mean-field BCS approximation, in an expectation that the topological characterization of pairing instabilities, will stay unchanged under this approximation. In the following we will review the relevant parts of the BCS mean-field theory. We will follow the notation of Ref. ~\onlinecite{rg}.
The
effective Hamiltonian is
\begin{equation}
K_{\rm eff} = \sum_{\bf k} \{ \xi_k a_{\bf k}^\dagger a_{\bf k} + \frac{1}{2} (\Delta^*_{\bf k} a_{- {\bf k}} a_{\bf k} + \Delta_{\bf k} a_{\bf k}^\dagger  a_{-{\bf k}}^\dagger) \},
\end{equation}
and in our case $\xi_k = E_k - \mu $, with $ E_k = \sqrt{|{\bf k}|^2 + m^2}$. The Bogoliubov transformation is
\begin{equation}
\alpha_{\bf k} = u_{\bf k} a_{\bf k} - v_{\bf k} a_{-{\bf k}}^{\dagger},
\end{equation}
with
\begin{eqnarray}
\frac{v_{\bf k}}{u_{\bf k}} & = & \frac{- ({\cal E}_k - \xi_k )}{\Delta_{\bf k}^{*}} , \nonumber \\
|u_{\bf k}|^2 & = & \frac{1}{2} (1 + \frac{\xi_k }{{\cal E}_k}), \nonumber \\
|v_{\bf k}|^2 & = & \frac{1}{2} (1 - \frac{\xi_k }{{\cal E}_k}),
\end{eqnarray}
and ${\cal E}_k = \sqrt{\xi^2_k + |\Delta_{\bf k}|^2}$.

On the other hand, if we start with a Cooper channel interaction and do the BCS mean field decomposition with
$ b_{\bf k}^\dagger = a_{\bf k}^\dagger a_{-{\bf k}}^\dagger $
\begin{eqnarray}
&& \sum_{{\bf k},{\bf p}} V_{{\bf k} {\bf p}} \; b_{\bf k}^\dagger \; b_{{\bf p}} = \sum_{{\bf k},{\bf p}} V_{{\bf k} {\bf p}} <b_{\bf k}^\dagger> b_{\bf p}  \nonumber \\
&& + \sum_{{\bf k},{\bf p}} V_{{\bf k} {\bf p}} b_{\bf k}^\dagger <b_{\bf p}>
- \sum_{{\bf k},{\bf p}} V_{{\bf k} {\bf p}} <b_{\bf k}^\dagger> <b_{\bf p}>, \nonumber \\
\end{eqnarray}
and specify $u_{- \bf k} = u_{\bf k} = u_{\bf k}^*$ and $v_{-{\bf k}} = - v_{{\bf k}}$,
then
\begin{eqnarray}
&&\frac{\Delta^*_{\bf p}}{2}  =  \sum_{\bf k} V_{{\bf k} {\bf p}} <a_{\bf k}^\dagger a_{-{\bf k}}^\dagger> \nonumber \\
&& =  \sum_{\bf k} V_{{\bf k} {\bf p}} <(u_{\bf k} \alpha_{\bf k}^\dagger + v_{\bf k}^* \alpha_{-{\bf k}}) (- v_{\bf k}^* \alpha_{{\bf k}} + u_{\bf k} \alpha_{-{\bf k}}^\dagger)>, \nonumber \\
\end{eqnarray}
i.e.
\begin{equation}
\frac{\Delta^*_{\bf p}}{2} = \sum_{\bf k} V_{{\bf k} {\bf p}} v_k^* u_k = \sum_{\bf k} V_{{\bf k} {\bf p}} (-) \frac{\Delta_{\bf k}^*}{2 \; {\cal E}_k}. \label{apsc}
\end{equation}

\begin{figure}[th]
\centerline{\includegraphics[width=13cm]{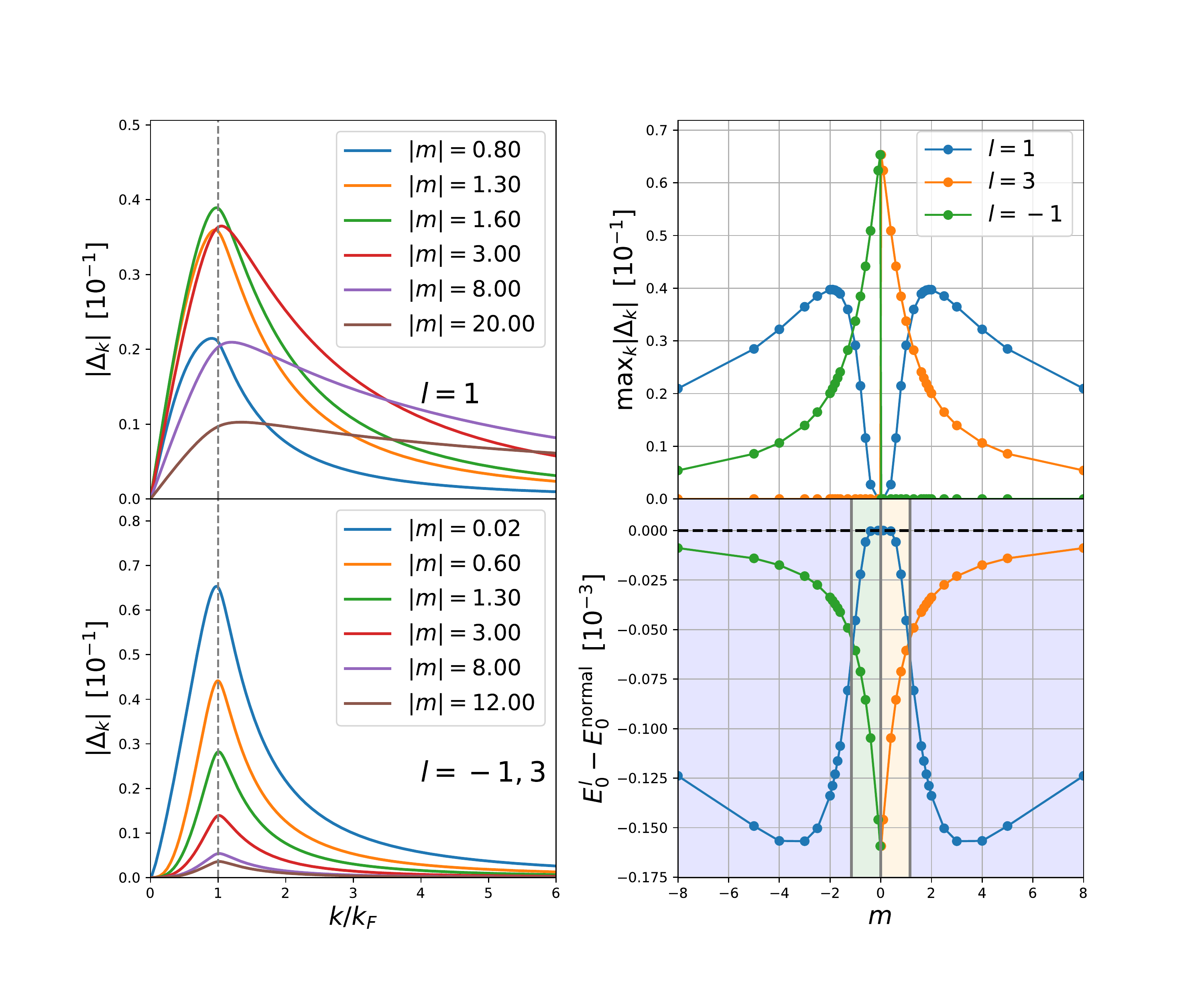}}
\vspace*{8pt}
%\begin{center}
%\begin{figure}[h!]
%\includegraphics[scale=0.6]{pfig.pdf}
\caption{The solution of the self-consistent BCS problem. Left column: radial direction $k$-dependent pairing amplitude for various values of $m$.
Channel $l = 1$ solution (PH Pfaffian) only depends on $|m|$, while $l = 3$ (anti-Pfaffian) and $l = -1$ (Pfaffian) channel solutions are symmetric with the sign-flip of $m$.
Upper right panel: dependence of the maximum of the pairing amplitude on $m$ (always found at the Fermi level $k_F$).
Lower right panel: total energy of the different pairing solutions compared to the normal state energy.
Gray vertical lines denote the transition between different channels. Color in the background corresponds to the energetiically favorable channel at the given $m$ - a measure of Landau level mixing. The color of lines: Pfaffian - green, anti-Pfaffian - orange, PH Pfaffian - blue.}
\end{figure}
%\end{center}

In our case  $V_{{\bf k}{\bf p}}$ is given in (\ref{Vkp}).  The numerical solutions of the BCS self-consistent equation, when the parameter $k_F$ is kept fixed,
but  mass $m$ is varied, for channels $l = 1, 3, -1$, with $\Delta_{\bf k}^{*} = |\Delta_{\bf k}| \exp\{i l \theta_{\bf k}\} $ are described in Fig. 1. We find
that $\Delta_{\bf k}^{*} = |\Delta_{\bf k}| \exp\{- i l \theta_{\bf k}\} $,  $l = 1, 3, -1$ are solutions if we switch gauge for the eigenstates of the Dirac equation, i.e.
instead of (\ref{eigenD})  we take
\begin{equation}
\chi_E = \left[
  \begin{array}{c}
  k_-       \\
   E_{\bf k} - m           \\
  \end{array}
\right]  \frac{1}{\sqrt{2 E_{\bf k} (E_{\bf k} - m)}}.
\end{equation}
Thus we get two sets of solutions, because the effective theory does not possess the knowledge of the direction of the external magnetic field. Despite this, we have a clear prediction that for small $m$ - LL mixing, depending on the sign of $m$ we have Pfaffian or anti-Pfaffian, and for large $m$  the PH Pfaffian solution is possible. Thus, in principle, the PH Pfaffian is possible in this effective theory of quasiparticle pairing. The nature of this state, whether it is gapped or gapless state of electrons, needs further investigations (though we see that the Bogoliubov quasiparticle spectrum is gapped). 

These predictions on topological pairing, when the LL mixing (mass $m$) is small, are in accordance with numerical experiments (a)  in the second LL,  because for $m = 0$ there is a Schroedinger cat superposition of Pfaffian and anti-Pfaffian \cite{rh1,dh}, and depending on the LL mixing (sign of PH breaking mass) we have Pfaffian or anti-Pfaffian, and (b) in the lowest LL,  where a PH Pfaffian wave function has a large overlap with the composite fermion Fermi-liquid wave function \cite{mish,bar}, in accordance with Fig. 1 where the PH Pfaffian-like state is continously connected to the excited composite fermion 
Fermi-liquid state at $m = 0$ and cannot represent a gapped state  in an isolated LL.

The dimensionless $m$ in the theory is a measure of the PH symmetry breaking and LL mixing, although the precise relation between $m$ and 
\begin{equation}
\kappa = \frac{\frac{e^2}{\epsilon_r l_B }}{\hbar \omega_c },
\label{kpp}
\end{equation}
 i.e. the ratio between the characteristic interaction energy and cyclotron energy, known as a
LL mixing coefficient, we do not know.  In (\ref{kpp}),  $\epsilon_r$ is the dielectric constant of the background material,  $ \hbar \omega_c = \frac{\hbar e B}{m_{\rm b} c} $, and $ m_{\rm b}$ is the electron band mass. As we keep the density, $\rho = \frac{\nu}{2 \pi l_B^2} = \frac{1}{2} \frac{1}{2 \pi l_B^2}$, i.e. $k_F$ fixed, from the mathematical limit of the PH symmetric case when $m = 0$, we reach various systems (experimental settings) by changing the interaction strength (dielectric constant $\epsilon_r$). Thus $m$, in principle, can be connected with $\kappa$, which can be considerable in  experiments.
 (According to Ref. \onlinecite{sm} the parameter $\kappa$ is given by $2.6/\sqrt{B}$, $14.6/\sqrt{B}$, $16.7/\sqrt{B}$, $22.5/\sqrt{B}$,  in n-doped GaAs, p-doped GaAs, n-doped ZnO, and n-doped AlAs, with $B$ measured in Tesla.)

\section{Model interactions for Pfaffian paired states}

It is important to know model interactions for model wave functions in order to probe their stability and nature.  In the case of bosons, the Pfaffian state at filling factor $1$ is
\begin{equation}
\Psi_{\rm Pf}^b  =   \sum_\sigma sgn \; \sigma \{ \frac{1}{(z_{\sigma(1)} - z_{\sigma(2)})}  \cdots
\frac{1}{(z_{\sigma(N_e -1)} - z_{\sigma(N_e )})}   \} \prod_{k<l} (z_{k} - z_{l}) .
 \label{bpf}
\end{equation}
The model interaction for which this state is an exact, densest state of zero energy \cite{gww} is
\begin{equation}
H = v \sum_{<ijk>} \delta^2 (z_i - z_j ) \delta^2 (z_i - z_k ) ,
\label{3bi}
\end{equation}
where $v > 0$ and the sum is over all distinct triples of particles. Thus if three bosons meet (come as close as possible) this will cost repulsive energy. In the case of fermions at filling factor 1/2, the Pfaffian model interaction is a generalization of the boson interaction to the one that,  if three fermions come as close as possible, again, only this will cost energy.  The lowest angular momentum wave function of three electrons in the lowest LL can be described as
\begin{equation}
\Psi({\bf r}_1 , {\bf r}_2 , {\bf r}_3 ) 
\sim
%=  \frac{1}{2^4 \pi^{3/2} l_B^6 \sqrt{3}}
 \sum_\sigma  sgn \; \sigma \; z_{\sigma(1)}^2  z_{\sigma(2)}^1  z_{\sigma(3)}^0
\; \exp\{ - \frac{1}{4 l_B} (|z_1 |^2  +
|z_2 |^2 + |z_3 |^2) \}.
\end{equation}
We may conclude that if $M ({\rm   angular \; momentum}) = 3$ for three electrons this will cost interaction energy. Indeed, it can be argued, 
just as in the case of the Laughlin state and two-body pseudo-potentials (PPs), \cite{hal} that in the case of Pfaffian we need to specify only a truncated series of three-body PPs with definite three-body angular momenta. At filling factor 1/2, only non-zero three-body PP is the one for $M = 3$. (For bosons, at filling factor 1, the only non-zero three-body PP is for $M = 0$.)

These model interactions are highly artificial if we want to model and probe real physical systems. In the FQHE we can always specify the base LL from which most of correlations originate, but should also consider the effects of LL mixing. Beside the Coulomb (two-body) interaction at a half-integer filling factor, we may take into account perturbatively the effects of LL mixing, by considering special three-body interactions. \cite{mix1, mix2, mix3, mix4, sm} In this way we may find  a characteristic series of three-body PPs for Pfaffian state, when considering the specific problem of the second LL and associated LL mixing contribution. A PP is a certain characteristic energy, $V_M$, associated with a  three-body state at total angular momentum $M$. (The dimension of the subspace of  a fixed angular momentum for three particles may be larger than one for higher $M$, and $V_M$ may be a matrix.) In the case of Pfaffian, the dominant, first three values of three-body PPs, for $M = 3, 5, 6$ are negative and $\frac{V_{M=5}}{V_{M=3}} \sim 0.4 $ and $\frac{V_{M=6}}{V_{M=3}} \sim 0.7 $. \cite{sm} We may ask what would be a characteristic series for PH Pfaffian, if we assume that the PH Pfaffian state or phase exists, and expect that some kind of three-body interaction will be relevant also in this case.

To answer this question we may consider again the Chern-Simons formalism, not directly connected with considerations in the previous section. We will recall  \cite{gww} the effective derivation of the Pfaffian physics, by a part of the kinetic term in the non-relativistic Chern-Simons description. (Thus these considerations will not relate to the solution in the previous section, in the large $m$ limit, when we take into account the complete kinetic term.) We will use this formal derivation to propose a method for recovering model interactions for Pfaffian and PH Pfaffian. (By using the particle-hole exchange we can reach a model interaction also for anti-Pfaffian.)

To get (formally) the Pfaffian pairing solution we may consider the kinetic energy part of the (non-relativistic) Chern-Simons approach in (\ref{hlrL}), i.e. the part of the Hamiltonian given by 
\begin{equation}
{\cal H} =  \frac{\Psi_{\rm cf}^{+} ({\bf p} - {\bf A} - {\bf a} )^2 \Psi_{\rm cf}}{2 m} ,
\label{hlrK}
\end{equation}
with $\bold{B} = B  \mathbf{z} $, with  $ A_{x} = - (B/2) y $ and $A_{y} = (B/2)  x $ , as before, and
\begin{equation}
a_x ({\bf r}) = 2   \int d{\bf r}' i \frac{y - y'}{|{\bf r} - {\bf r}'|^2} \delta \rho_{\rm cf} ({\bf r}'), \label{axp}
\end{equation}
and
\begin{equation}
a_y  ({\bf r}) = -  2   \int d{\bf r}' i \frac{x - x'}{|{\bf r} - {\bf r}'|^2} \delta \rho_{\rm cf} ({\bf r}'), \label{ayp}
\end{equation}
as before,  in the Coulomb gauge $ {\bf \nabla} \cdot {\bf a} = 0$, and  $\delta \rho_{\rm cf} = \Psi_{\rm cf}^{+} \Psi_{\rm cf} - \overline{\rho}$, where $\overline{\rho}$ is the average density.    We  consider the following part of the implied interaction,
\begin{equation}
V_{\rm a} = - {\bf a} {\bf j}_{\rm cf},
\label{pint}
\end{equation}
with
\begin{equation}
{\bf j}_{\rm cf} = \frac{1}{2 m} [ \Psi^+_{\rm cf} ({\bf p} \Psi_{\rm cf}) - ({\bf p} \Psi^+_{\rm cf})  \Psi_{\rm cf}],
\end{equation}
more specifically its Cooper channel part. 

After simple steps  \cite{djm} we arrive at the Cooper channel part,
\begin{equation}
V_{\rm int}^{\rm C}  =   \frac{4 \pi }{m} \frac{1}{V} \sum_{\mathbf{k}, \mathbf{p}} |{\bf k}| |{\bf p}|  \frac{ i \sin (\theta_k - \theta_p )}{|\mathbf{p} - \mathbf{k}|^2}    a^\dagger_{\mathbf{k}} 
a_{\mathbf{p}}   a^\dagger_{-\mathbf{k}} 
a_{-\mathbf{p}} .
\label{cchp}
\end{equation}
Note that in this case (following the mean field equations and derivation in Ref.~\onlinecite{rg}, or in Ref.~\onlinecite{djm}) we find that the Cooper pair wave function behaves as,
\begin{equation}
\lim_{|{\bf r}| \rightarrow \infty} g({\rm r}) \sim \frac{1}{z}.
\end{equation}
This implies the Pfaffian construction (after the unitary Chern-Simons transformation into the electron representation), if we recall that the choice of ${\bf A}$ in (\ref{hlrK}) implies a holomorphic Laughlin-Jastrow factor (more precisely a phase factor after the unitary Chern-Simons transformation) that is associated
with the usual description of the Pfaffian state in (\ref{pf}). If we had an extra minus sign in (\ref{cchp}) this would lead to the anti-holomorphic pairing, i.e. 
the PH Pfaffian pairing.

To derive the model interactions for Pfaffian and PH Pfaffian we assume that we can use an effective non-relativistic Chern-Simons description to describe the pairing of underlying qusiparticles (composite fermions). On the basis of the previous consideration ((\ref{pint})  and (\ref{cchp})), we consider an effective Hamiltonian,
\begin{equation}
H_{\rm BCS}^{\rm ef} = \frac{1}{2 m} \Psi_{\rm cf}^\dagger (\mathbf{p})^2 \Psi_{\rm cf} + \lambda \delta \mathbf{a} \mathbf{j}_{\rm cf} ,
\label{bHam}
\end{equation}
where $\delta \mathbf{a} = \mathbf{A} + \mathbf{a}$, and the coupling $\lambda$ is negative in the Pfaffian case and positive in the PH Pfaffian case.
Thus we assumed that a complete (non-relativistic) Chern-Simons description that includes all effects of interactions can be reduced to the effective form
if a pairing occurs. By using the non-relativistic Chern-Simons description we take into account particle-hole symmetry breaking necessary to stabilize these 
pairing states.

If we apply the Chern-Simons transformation in reverse,  \cite{djm} going from the composite fermion representation to an electron one,  we arrive at the following effective Hamiltonian for electrons,
\begin{equation}
H_{\rm BCS}^{\rm el} = \frac{1}{2 m} \Psi^\dagger (\mathbf{p} -  \mathbf{A})^2 \Psi - \frac{1}{2 m} (\delta \mathbf{a} )^2 \Psi^{\dagger} \Psi  + (1 + \lambda) \delta \mathbf{a} \mathbf{J}_{\rm el}  + (1 + \lambda ) \frac{1}{m}  (\delta \mathbf{a} )^2 \Psi^{\dagger} \Psi  ,
\label{Ham}
\end{equation}
where
\begin{equation}
\mathbf{J}_{\rm el} =  \frac{ - i}{2 m} { \Psi^\dagger (\mathbf{\nabla}  + i \mathbf{A} ) \Psi  - [ (\mathbf{\nabla}  + i \mathbf{A} ) \Psi ]^\dagger  \Psi },
\label{gc}
\end{equation}
is the (gauge invariant) electron current.

We concentrate on the effective three-body (electron) interaction that is present in the Hamiltonian,
\begin{equation}
V_{\rm BCS}^{3}(\lambda) =   (1/2 + \lambda ) \frac{1}{m} : ( \mathbf{a} )^2 \Psi^{\dagger} \Psi :  .
\label{Int3}
\end{equation}
The three-body interaction in coordinate representation is
\begin{equation}
V (\mathbf{r}_1 , \mathbf{r}_2 , \mathbf{r}_3 ) =   (1/2 + \lambda ) \frac{4}{m}    \frac{(\mathbf{r}_3 - \mathbf{r}_1 )  (\mathbf{r}_3 - \mathbf{r}_2 )}{|\mathbf{r}_3 - \mathbf{r}_1 |^{2} |\mathbf{r}_3 - \mathbf{r}_2 |^{2}}. \label{3V}
\end{equation}

To describe the relevant matrix elements for LL(s) we will choose our base LL to be the lowest LL, which is the most natural choice when we consider a Chern-Simons description; the very Chern-Simons transformation is based on the Laughlin-Jastrow correlations in the lowest LL. Thus, for example, we will relate the effective PPs that we know for the Pfaffian state, based on the perturbation theory, in the second LL, with here calculated PPs, based on the Chern-Simons description, in the lowest LL.

To describe relevant three-body PPs $(V_M )$ in the lowest LL, we introduce rescaled matrix elements, $\Delta_{M = 2 k + 3 l}$,
\begin{eqnarray}
V_M  = \int d \mathbf{r}_1 \int d \mathbf{r}_2 \int d \mathbf{r}_3
 V (\mathbf{r}_1 , \mathbf{r}_2 , \mathbf{r}_3 )  | \Psi_{k,l} (\mathbf{r}_1 , \mathbf{r}_2 , \mathbf{r}_3 ) |^2  \nonumber
= (1/2 + \lambda ) \cdot 4/m  \cdot \Delta_{M = 2 k + 3 l} ,
\end{eqnarray}
where $\Psi_{k,l}$ are normalized, fully antisymmetric wave functions for three electrons,  \cite{laugh} classified by integers  $ k \geq 0 ; \;  l \geq 1 $, and
the total angular momentum of the state is $ M = (2 k + 3 l ) $.
The caculated $\Delta_M$ are shown in the Table I.

\begin{table}[pt]
%\begin{table}[h!]
%\begin{center} 
\caption{Matrix elements in the lowest Landau level.}
\begin{tabular}{|c||c|c|c|c|c|c|}
\hline
M & 3 & 5 & 6 & 7 & 8 & 9  \\
\hline\hline
$\Delta_M$ & $1/24$ & $1/48$ & $7/240$ & $1/80$ & $2/105$ & 
\begin{tabular}{l|l}
$221/10080$ & $1/(240\sqrt{21})$\\
\hline
$1/(240\sqrt{21})$ & $1/120$
\end{tabular}  \\
\hline
$\frac{\Delta_M}{\Delta_{M=3}}$ & $1$ & $0.5$ & $0.7$ & $0.3$ & $\sim 0.475$ & 
\begin{tabular}{l|l}
$\sim 0.526$ & $\sim 0.022$\\
\hline
$\sim 0.022$ & $0.2$
\end{tabular}  \\
\hline
\end{tabular}
%\end{center}
\end{table}

%\begin{table}[pt]
%\begin{center} 
%\caption{Matrix elements in the lowest Landau level.}
%\begin{tabular}{|c||c|c|c|c|c|c|c|}
%\hline
%M & 3 & 5 & 6 & 7 & 8 & 9 & 10 \\
%\hline\hline
%$\Delta_M$ & $1/24$ & $1/48$ & $7/240$ & $1/80$ & $2/105$ & 
%\begin{tabular}{l|l}
%$221/10080$ & $1/(240\sqrt{21})$\\
%\hline
%$1/(240\sqrt{21})$ & $1/120$
%\end{tabular} & $3/224$ \\
%\hline
%$\frac{\Delta_M}{\Delta_{M=3}}$ & $1$ & $0.5$ & $0.7$ & $0.3$ & $\sim 0.475$ & 
%\begin{tabular}{l|l}
%$\sim 0.526$ & $\sim 0.022$\\
%\hline
%$\sim 0.022$ & $0.2$
%\end{tabular} & $\sim 0.321$ \\
%\hline
%\end{tabular}
%\end{center}
%\end{table}

\begin{figure}[th]
\centerline{\includegraphics[width=8cm]{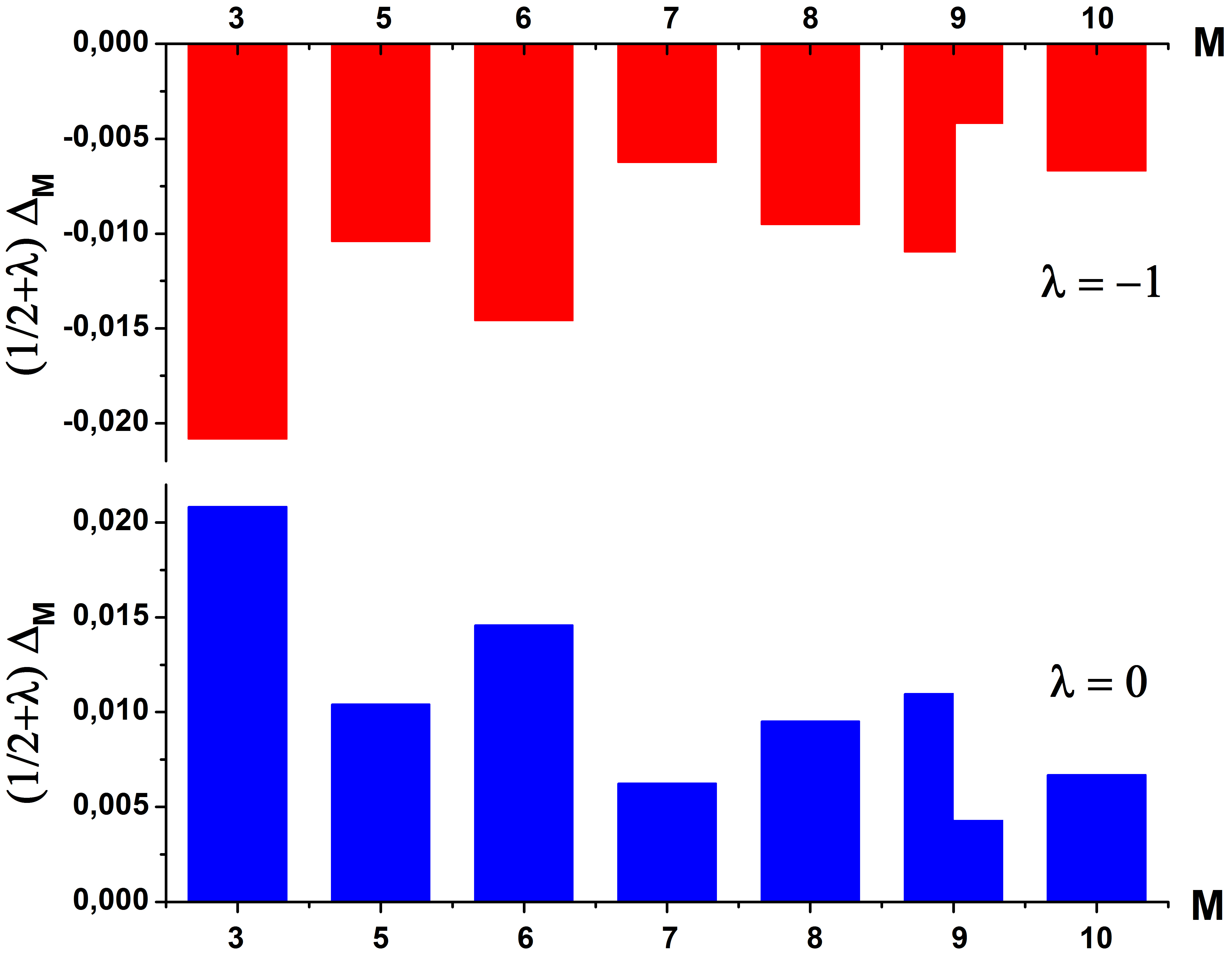}}
\vspace*{8pt}
%\begin{center}
%\begin{figure}[h!]
%\includegraphics[scale=0.4]{slikica11.jpg}
\caption{Matrix elements of three body pseudo-potentials in the lowest Landau level for $\lambda=-1$ (above) and $\lambda=0$ (bottom).  (We plotted two values, diagonal matrix elements in the two-dimensional subspace, in the case when $M = 9$.)}
\end{figure}
%\end{center}
The matrix elements are illustrated by their rescaled values $ \frac{m}{4} V_M = (1/2 + \lambda )   \cdot \Delta_{M = 2 k + 3 l} $, in the cases when
$\lambda = -1$ and $\lambda = 0$ in Fig. 2. What is remarkable is that according to the Table, $\frac{V_{M=5}}{V_{M=3}} = 0.5 $, and  $\frac{V_{M=6}}{V_{M=3}} = 0.7 $,  and are quite close to the ratios of the relevant matrix elements from the perturbation theory in the second LL, $ \sim 0.4$, and $\sim 0.7 $, respectively,
that favor the Pfaffian physics. \cite{pp}

Thus the Chern-Simons description is able to capture the sign - a negative one of necessary PPs when $\lambda < - 1/2$, and their relative magnitude for
relevant, those first three PPs in the Pfaffian case. Therefore we are encouraged to probe the PH Pfaffian case for certainly $\lambda > 0$. (We can identify the $\lambda = 0$ case with composite fermion Fermi liquid case.) But we have to be aware that in the effective description by $H_{\rm BCS}^{\rm el}$, the estimate that we can make for LL mixing parameter (in general the ratio of characteristic interaction energy and cyclotron energy) is $|\lambda +1/2|$, and that for any considerable $\lambda  \gtrsim  1/2$ for which PH Pfaffian correlations are relevant, we have to include higher LL(s) (i.e. not only the base LL - the lowest LL in the Chern-Simons description).

Thus in the PH Pfaffian case we have to include (three-body) PPs for at least one more LL. The calculated PPs (more precisely their rescaled  $ (m/4) V_M$ values) for two LLs when $\lambda = 1$ are illustrated in Fig. 3.
\begin{figure}[th]
\centerline{\includegraphics[width=8cm]{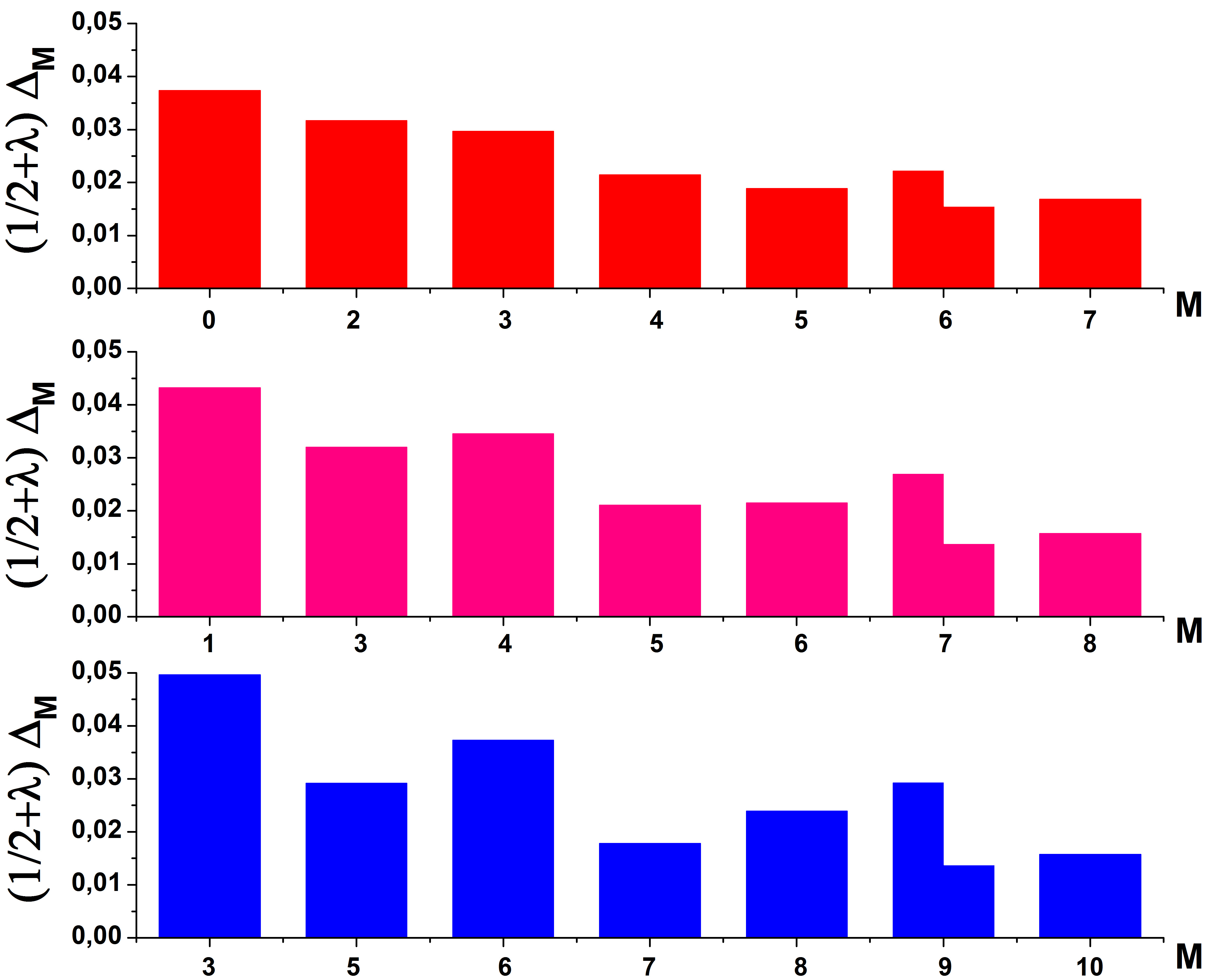}}
\vspace*{8pt}
%\begin{center}
%\begin{figure}[h!]
%\includegraphics[scale=0.4]{slikica22.jpg}
\caption{Three body pseudo-potential matrix elements for $\lambda=1$  (PH Pfaffian case)  in the second Landau level (top), for states with two particles in the second Landau level and one in the lowest Landau level (middle), and (all three) in the lowest Landau level (bottom).}
\end{figure}
%\end{center}
While calculating these PPs we had to include the natural cut-off $l_B$ in the field theoretical description, to suppress divergences in the second LL. We can conclude from Fig. 3 that in the case of PH Pfaffian, there is an abrupt decrease in the positive values of three-body PPs at $M = 7$ in the base (lowest LL) level and also at $M = 5$, when two of three electrons are in the higher (second) LL. This can be compared with the usual (truncated) model for Pfaffian with only non-zero, positive potential $V_{M=3}$; there is no three fermion state with $M = 4$, and the $V_5$ PP that is connected with the characteristic  three-body angular momentum for Pfaffian in the LLL, $M = 5$,  is zero. \cite{src} In the case of the PH Pfaffian the characteristic angular momentum is $M = 7$ in the lowest LL, and thus the abrupt decrease(s)  in the values of three-body PPs that we may associate with the PH Pfaffian pairing correlations.  The (almost) monotonic decrease of PPs when all three particles are in the second LL suggests that the space of two LLs may be necessary, but also sufficient for the realization of the PH Pfafian correlations. The important question, which needs further investigation, is whether these correlations are associated with a gapped state. The 
most recent suggestion for the realization of  PH Pfaffian is in Ref.~\onlinecite{phdf}.

\section{Conclusions and outlook}

In this review we have demonstrated that the Chern-Simons field-theoretical approach can be useful and informative in the description of Pfaffian and anti-Pfaffian states - well-established candidate states for the explanation of gapped states at half-integer filling factors in the FQHE. It can capture the pairing nature of these states, when the basic gauge-field constraints are taken into account in a generalized Dirac effective description of the problem. The effective Dirac description originates from the physics inside a base LL, which, when isolated (in the case of the Coulomb problem) possesses PH symmetry. To stabilize Pfaffian or anti-Pfaffian we have to break this symmetry by a mass (of definite sign) term in the Dirac theory.

The physics of an isolated base LL in the Dirac effective description suggests a possible existence of a PH symmetric Pfaffian state.  \cite{son} We find that this solution is relevant only when a significant PH breaking (mass) is included in the Dirac description. Considering a non-relativistic limit of the description we find that interaction parameters that describe the influence  from the higher (second) LL must be  nonperturbatively included in a model interaction for PH Pfaffian (beside the ones from the base (lowest) LL). This may be helpful in the effort to stabilize and detect PH Pfaffian correlations in numerical experiments.

\section*{Acknowledgments}

  This research
was supported by the Ministry of Education, Science, and
Technological Development of the Republic of Serbia under
Project ON171017, and by the Ministry of Science of Montenegro under Project SFS013454.

%This section should come before the References. Funding
%information may also be included here.

%\appendix

%\section{Heading of Appendix}

%Appendices should be used only when absolutely necessary. They
%should come after the References. If there is more than one
%appendix, number them alphabetically, e.g.~Appendix~A, Appendix~B, etc.
%Number displayed equations
%occurring in the Appendix in this way, e.g.~(\ref{that}), (A.2),
%etc.

%\noindent
%\begin{equation}
%\mu(n, t) = \frac{\sum^\infty_{i=1} 1(d_i < t,
%N(d_i) = n)}{\int^t_{\sigma=0} 1(N(\sigma) = n)d\sigma}\,. \label{that}
%\end{equation}

\end{widetext}

\end{document}